\documentclass[]{aastex6}

\usepackage{graphicx} 
\usepackage{bm} 
\usepackage{hyperref} 

\begin{document}


\title{
Search for Neutrinos in Super-Kamiokande associated with the GW170817 neutron-star merger
}





\keywords{astroparticle physics --- gravitational waves --- neutrinos}

\author{
K.~Abe\altaffilmark{1,37},
C.~Bronner\altaffilmark{1},
Y.~Hayato\altaffilmark{1,37},
M.~Ikeda\altaffilmark{1},
K.~Iyogi\altaffilmark{1},
J.~Kameda\altaffilmark{1,37},
Y.~Kato\altaffilmark{1,37},
Y.~Kishimoto\altaffilmark{1,37},
Ll.~Marti\altaffilmark{1},
M.~Miura\altaffilmark{1,37}
% ...
}
\altaffiltext{1}{Kamioka Observatory, Institute for Cosmic Ray Research, University of Tokyo, Kamioka, Gifu 506-1205, Japan}
% ...
\altaffiltext{37}{STFC, Rutherford Appleton Laboratory, Harwell Oxford, and Daresbury Laboratory, Warrington, OX11 0QX, United Kingdom}
% ...


\collaboration{The Super-Kamiokande Collaboration}
\noaffiliation

\begin{abstract}
We report the results of a neutrino search in Super-Kamiokande for coincident signals with the first detected gravitational wave produced by a binary neutron star merger, GW170817, which was followed by a short gamma-ray burst, GRB170817A, and a kilonova/macronova.
We searched for coincident neutrino events in the range from 3.5~MeV to 
$\sim$100~PeV, in a time window $\pm$500 seconds around the gravitational wave detection time, as well as during a 14-day period after the detection.
No significant neutrino signal was observed for either time window.
We calculated 90\% confidence level upper limits on the neutrino fluence for GW170817.
From the upward-going-muon events in the energy region above 1.6~GeV, the neutrino fluence limit is $16.0^{+0.7}_{-0.6}$~($21.3^{+1.1}_{-0.8}$) cm$^{-2}$ for muon neutrinos (muon antineutrinos), with an error range of $\pm5^{\circ}$ around the zenith angle of NGC4993, and the energy spectrum is under the assumption of an index of $-2$.
The fluence limit for neutrino energies less than 100~MeV, for which the emission mechanism would be different than for higher-energy neutrinos, is also calculated.
It is 6.6 $\times 10^7$ cm$^{-2}$ for anti-electron neutrinos under the assumption of a Fermi-Dirac spectrum with average energy of 20~MeV.
\end{abstract}

\renewcommand{\thefootnote}{\fnsymbol{footnote}}
\footnote[0]{$^{\dagger}$ Also at Department of Physics and Astronomy, UCLA, CA 90095-1547, USA.}
\footnote[0]{$^{\ddagger}$ Also at BMCC/CUNY, Science Department, New York, New York, USA.}



\section{Introduction} \label{sec:introduction}
On August 17th 2017 at 12:41:04 UTC, the Advanced LIGO and Advanced Virgo experiment identified the first evident signal of a gravitational wave from the binary neutron star merger, named GW170817~\cite{2017PhRvL.119p1101A}.
The interpretation is a merger of two compact objects consistent with neutron stars having total system mass of 2.74 solar masses and a luminosity distance of 40 Mpc.
Associated with this gravitational wave signal, the Fermi Gamma-ray Burst Monitor and International the Gamma-Ray Astrophysics Laboratory also detected a short gamma-ray burst, GRB170817A, which has a consistent location with the merger and a 1.7-s delay to the merger time~\cite{2017ApJ...848L..13A}.
Subsequent extensive electromagnetic follow-up observations in ultra-violet, optical and infrared wavelengths were performed.
These observations led to the conclusion that the merger happened in galaxy NGC4993 and was followed by a short gamma-ray burst and a kilonova/macronova~\cite{2017ApJ...848L..12A}.
High-energy neutrino signals associated with the merger were also searched for by the ANTARES, IceCube, and Pierre Auger Observatories.
It was concluded that no significant neutrino signal was observed~\cite{2017ApJ...850L..35A}.

We report the results of a search for neutrinos in Super-Kamiokande (SK) associated with this gravitational wave signal produced by the binary neutron star merger in NGC4993.
The analysis method is similar to that for the previous neutrino search in SK for GW150914 and GW151226~\cite{2016ApJ...830L..11A}.
SK is a water Cherenkov detector with 50-kton water mass and 22.5-kton fiducial volume. It is located 2700-meters-water-equivalent underground in Kamioka, Japan.
A detailed description of the detector, its calibration and performance can be found elsewhere~\cite{2003NIMPA.501..418F, 2014NIMPA.737..253A}.
In this detector, the Cherenkov ring pattern reconstruction identifies final-state electron and muon direction and energy, from which we infer the neutrino direction, flavor, and energy.
SK has sensitivity to  a wide neutrino energy region and is able to reconstruct neutrino event energies from a few MeV to $\sim$100~PeV.
Neutrino events with reconstructed energies above 100~MeV are categorized as the `high-energy data sample' in SK and are typically used to study atmospheric neutrinos and to search for proton decay.
Neutrino events with reconstructed energies down to 3.5~MeV are categorized as the `low-energy data sample' and are typically used to study solar neutrinos and to search for core-collapse supernova neutrinos.
The directional determination accuracy varies according to sample and direction, but can be as accurate as $\sim$1 degree for upward-going muons.
Some theoretical predictions of neutrino emission mechanism via binary neutron star mergers have been proposed; for example, some fraction of the kinetic energy in relativistic ejecta from gamma-ray bursts could convert to high-energy ($\sim10^{14}$~eV) neutrinos~\cite{1997PhRvL..78.2292W}, or a similar mechanism as for core-collapse supernovae could produce few-tens-of-MeV  neutrinos~\cite{2011PhRvL.107e1102S, 1710.05922}, the expected fluence is roughly estimated to be $10^4$ cm$^{-2}$ for 10 MeV neutrinos within 1 second after merger.
Neutrino observations associated with a binary neutron star merger using the unique characteristics in SK would validate such proposed mechanisms.
We searched for coincident events in the full data sample using the same time window as ANTARES-IceCube-Pierre Auger, i.e., $\pm$500 s around the merger time and in a 14-day time window relevant for longer-lived emission processes.
The primary background events for this search in the high-energy data sample are almost entirely atmospheric neutrinos, while radioactive impurities, spallation products from cosmic ray muons, atmospheric and solar neutrinos are the main backgrounds in the low-energy data sample.
We note that SK carried out a LINAC calibration~\cite{1999NIMPA.421..113N} from August 3-22, 2017.
Fortunately, physics data-taking operated at the time when the neutron star merger occurred; however, there were unavoidable radioactive impurities adhered on the surface of the LINAC beam pipe present in the low-energy data sample.

\section{Search method and results}\label{sec:searchmethod}
\subsection{High-energy data sample}

The high-energy data sample has three different categories: fully-contained (FC), partially-contained (PC), and upward-going muon (UPMU). 
FC neutrinos have reconstructed interaction vertices inside the fiducial volume of the inner detector, combined with low light levels in the outer detector. 
PC neutrinos also have interaction vertices inside the fiducial volume, but have significant light in the outer detector volume indicating exiting particles. 
UPMU neutrinos are the highest-energy SK sample; they result from muon-neutrino interactions in the rock surrounding the detector, which produce penetrating muons. These muons either stop in the inner-detector volume as stopping events, or go through the inner detector as through-going events. 
The energy range for neutrino parents in FC and PC sample is 100 MeV--10 GeV, and for UPMU it is 1.6 GeV--100 PeV. All the three event topologies are considered for this search. 
Further information about the selection and reconstruction methods for the three categories can be found in \cite{ashie05}. 

\begin{table}[hptb]
\begin{center}
\begin{tabular}{lc c c}\hline \hline
 & observed num. of event  & expected num. of event\\ \hline
 & {\bf in $\pm$500 s} \\
FC       & 0  &  $(9.36\pm0.06)\times 10^{-2}$ \\
PC       & 0  &  $(7.52\pm0.23)\times 10^{-3}$ \\
UPMU     & 0  &  $(1.64\pm0.02)\times 10^{-2}$  \\
\hline
 & {\bf following 14 days for all sky} \\
FC       & $76\pm8.72$  &  $91.44\pm0.57 $ \\
PC       & $8\pm2.83$  &  $7.35\pm0.23 $  \\
 & {\bf following 14 days for $5^{\circ}$ solid angle} \\
UPMU      & 0  &  $(6.11\pm0.04) \times 10^{-2} $ \\

\hline
\end{tabular}
\caption{The numbers of expected and observed events in FC, PC and UPMU data sets, respectively, for a $\pm$500-s time window around GW170817 and for 14 days after GW170817. The errors on the observed number of events in the following 14 days for the entire sky are $\sqrt{N}$.
For UPMU event search in following 14 days, the event number in a solid angle of $\pm5^{\circ}$ around NGC4993 was shown instead. 
The total livetime for the following 14 days is 11.30 days.}
\label{event_table}
\end{center}
\end{table}

A $\pm$500-s window search around the LIGO detection time of GW170817, as well as a 14-day window search following the GW detection, have been conducted in the SK detector. These windows are consistent with those selected in \cite{2017ApJ...850L..35A}. 
The expected number of events based on 2976.01 days of SK data, and the number of events we actually observed, are listed in Table~\ref{event_table} for a $\pm$500-s window and for the following 14-day interval. 
The livetime for the high-energy analysis after removing LINAC beam runs for the following 14 days after GW170817 is 11.30 days. 
In a $\pm$500-s window around GW170817, no neutrino event was found in the FC, PC, or UPMU data sets. This null result is used in the calculation of the upper limit on neutrino fluence in the subsequent sections of this letter.
Unlike the FC and PC samples, the UPMU sample only contains upgoing muons, so it is sensitive to only half of the sky. In 60.4\% of the following 14 days, NGC4993 is within the sensitive half. Since the direction of NGC4993 is well known~\cite{2017ApJ...848L..12A}, for UPMU data, for which the angular resolution is better than for the other two samples, we concentrated on a $\pm5^{\circ}$ cone around NGC4993 for the event search in the following 14 days.
This method was previously used in SK to search for neutrino signals associated with astrophysical objects \cite{erin_icecube}. The $5^{\circ}$ constraint was not used for the $\pm$500-s search in Table~\ref{event_table} because no event was observed in all sky during this window, and unlike the 14-day-window case, the zenith angle change of NGC4993 in $\pm$500-s can be ignored.
All the results listed in Table~\ref{event_table} are consistent with our expected event rates.

\subsection{Low-energy data sample}
Assuming the flux is approximately equally distributed among flavors, as for a core-collapse supernova, the dominant channel in the 3.5 MeV--100 MeV range is the inverse beta decay of electron antineutrinos ($\bar{\nu}_e+p \rightarrow e^+ + n$). 
The second most dominant one is neutrino elastic scattering ($\nu + e^{-} \rightarrow \nu + e^{-}$), which is sensitive to all neutrino flavors, but dominated by electron neutrinos. 
Positrons or electrons from these interactions can produce observable signals in the SK detector. There are other charged-current and neutral-current interactions with $^{16}$O nuclei which are subdominant.

There are two data samples used by SK for low-energy analysis: one tuned for the supernova relic neutrino (SRN) search~\cite{sksrn} and another for the solar neutrino analysis~\cite{sk4sol}. These are independent selection methods and apply to different energy ranges for the GW170817 event search.
The solar neutrino analysis is applied in the 3.5 MeV--15.5 MeV range, while the SRN analysis focuses on the 15.5 MeV--100 MeV range.
In the low-energy analysis, the main background under 20 MeV is spallation products from cosmic-ray muons, and above 20 MeV the dominant background is from atmospheric-neutrino interactions (decay electrons from invisible muons, neutral-current interactions, low-energy pions and muons).
It should be mentioned that, in this analysis for the GW170817 event search, we do not require a neutron signal~\cite{ntag}. 

After all reduction steps, no neutrino was observed in the SRN analysis within the $\pm500$-s search window around GW170817. The expected number of background events in a $\pm$500-s time window is $(1.93\pm0.08)\times 10^{-3}$, based on 2887 days of data.

In the SRN sample for the following 14 days after GW170817, two events were found, on August 24th 10:33:04 UTC and on August 28th 14:36:34 UTC. The reconstructed energies of these two events are 22.0 MeV and 40.4 MeV in kinetic energy, and the angles between their reconstructed directions and NGC4993 are $(55.41\pm15.81)^{\circ}$ and $(145.24\pm11.30)^{\circ}$. 
Because of the LINAC calibration, livetime for the following 14 days of SRN-analysis data is 9.15 days and the expected number of events is $1.53\pm0.06$, so the probability of observing two or more events is 45.1\%. 
This is consistent with the expected signal rate and we do not classify these as GW170817 neutrinos rather than SRN candidates. 
Therefore, the fluence limit calculation, which will be discussed in next section, is performed based on the result of the $\pm$500-s window search.

The same search windows were applied to the solar neutrino data and no event was found in a $\pm$500-s window around GW170817. 
Using the data after May 1, 2015, whose livetime is 306.6 days, we expect $2.90\pm0.01$ events in $\pm$500 s and the probability of no event is 5.5\%. Due to LINAC calibration work during August 2017, tank opening and hardware changes were carried out so frequently that the quality of the event selection and reconstruction in the solar neutrino data sample was not stable, primarily due to radioactive impurities from the LINAC pipe. Therefore we omit discussion of the following 14-day data sample.

\section{Neutrino fluence limit} \label{sec:fluence}
As there is no event observed within a $\pm$500-s window, either in low-energy data nor in high-energy data, the null number can be converted to an upper limit on neutrino fluence. This is done separately for the low-energy, FC+PC, and UPMU data sets. The fluence limit was calculated using the same procedure laid out in \cite{thrane09}, which follows from \cite{swanson06}. 

For the FC and PC data set, the neutrino fluence can be calculated using equation (\ref{eqn:fluence}), 

\begin{equation} \label{eqn:fluence}
\Phi_{FC, PC} = \frac{N_{90}}{N_T \int dE_{\nu} \sigma (E_{\nu}) \epsilon (E_{\nu}) \lambda (E^{-2}_{\nu})},
\end{equation}

$N_{90}$ is the 90\% C.L. limit calculated from a Poisson distribution, for the observed neutrino events in a $\pm500$-s window with the expected background number. 
Since there is no neutrino event found in a $\pm500$-s window for FC, PC and UPMU data, $N_{90}$ can be fixed as $N_{90} = - \ln(0.1) =$ 2.3.
$N_{T}$ is the number of target nuclei relevant to the neutrino interactions. $\sigma$ is the combined cross section for all interactions. $\epsilon$ is detection efficiency, and $\lambda$ is the density of $E_{\nu}$ assuming an energy spectrum with index of $-2$. This spectral index is commonly assumed for astrophysical neutrinos accelerated by shocks \cite{gaisser94}. 

Fluence limits are calculated separately for each neutrino type because the cross section and detection efficiency depend on neutrino type.
Cross sections in equation (\ref{eqn:fluence}) are from NEUT 5.3.5 \cite{hayato09}. NEUT 5.3.5 is also used to produce mono-energetic neutrino interactions in the SK Monte Carlo detector simulation in order to determine the detection efficiency. 

For the UPMU data set, the neutrino fluence is calculated using
equation (\ref{eqn:UPMUfluence}),
 
\begin{equation} \label{eqn:UPMUfluence}
\Phi_{UPMU} = \frac{N_{90}}{A_{eff}(z)\int dE_{\nu} P(E_{\nu})S(z,E_{\nu})\lambda (E^{-2}_{\nu})}.
\end{equation}

The fluence of UPMU events depends on zenith angle. $A_{eff}(z)$ is the zenith-dependent effective area, where $z$ is the the zenith angle of the incoming neutrino. 
$P(E_{\nu})$ is the probability for a neutrino to create a muon with energy greater than $E_{\nu}^{min}$.
$S(z,E_{\nu})$ is the shadowing of the neutrinos due to interactions in
the Earth.
As for the FC and PC analysis, $\lambda$ here is the number density of $E_{\nu}$ in a spectrum with index of $-2$.

The fluence calculation for low-energy neutrinos uses an expression similar to (\ref{eqn:fluence}) but with different energy spectra. Here we assume two kind of spectra, one with an index of $0$, i.e.,  a flat spectrum, and another being a Fermi-Dirac distribution with average energy of 20 MeV:

\begin{equation} \label{eqn:lowe}
\Phi_{lowe} = \frac{N_{90}}{N_T \int dE_{\nu} \lambda (E_{\nu}) \sigma (E_{\nu}) R(E_e, E_{vis}) \epsilon (E_{vis}) },
\end{equation}

$R$ is the response function to convert electron or positron energy ($E_{e}$) to kinetic energy in SK ($E_{vis}$). The response function and the detection efficiency ($\epsilon$) are calculated using SK detector Monte Carlo simulation. 
Again, no event was observed in $\pm$500~s, so $N_{90}$ is 2.3. 

\begin{figure}[hptb]
\begin{center}
\includegraphics[width=10cm]{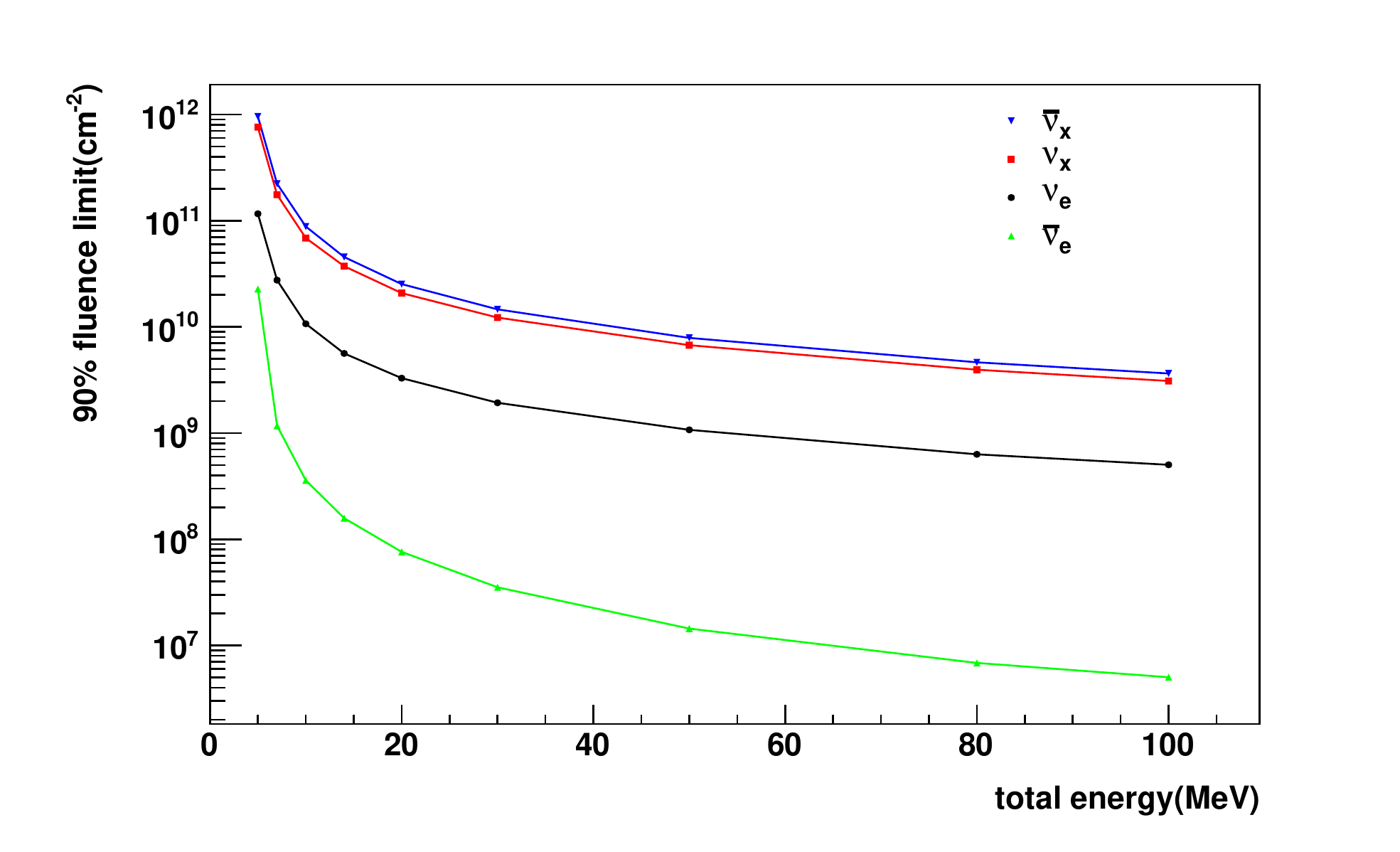}
\end{center}
\caption{The 90\% C.L. limits for GW170817 events on fluence obtained for mono-energetic neutrinos at 4 MeV, 7 MeV, 10 MeV, 14 MeV, 20 MeV, 30 MeV, 50 MeV, 80 MeV and 100 MeV.}
\label{fig:lowe_g}
\end{figure}

\begin{figure}[hptb]
\centering
\includegraphics[width=8cm]{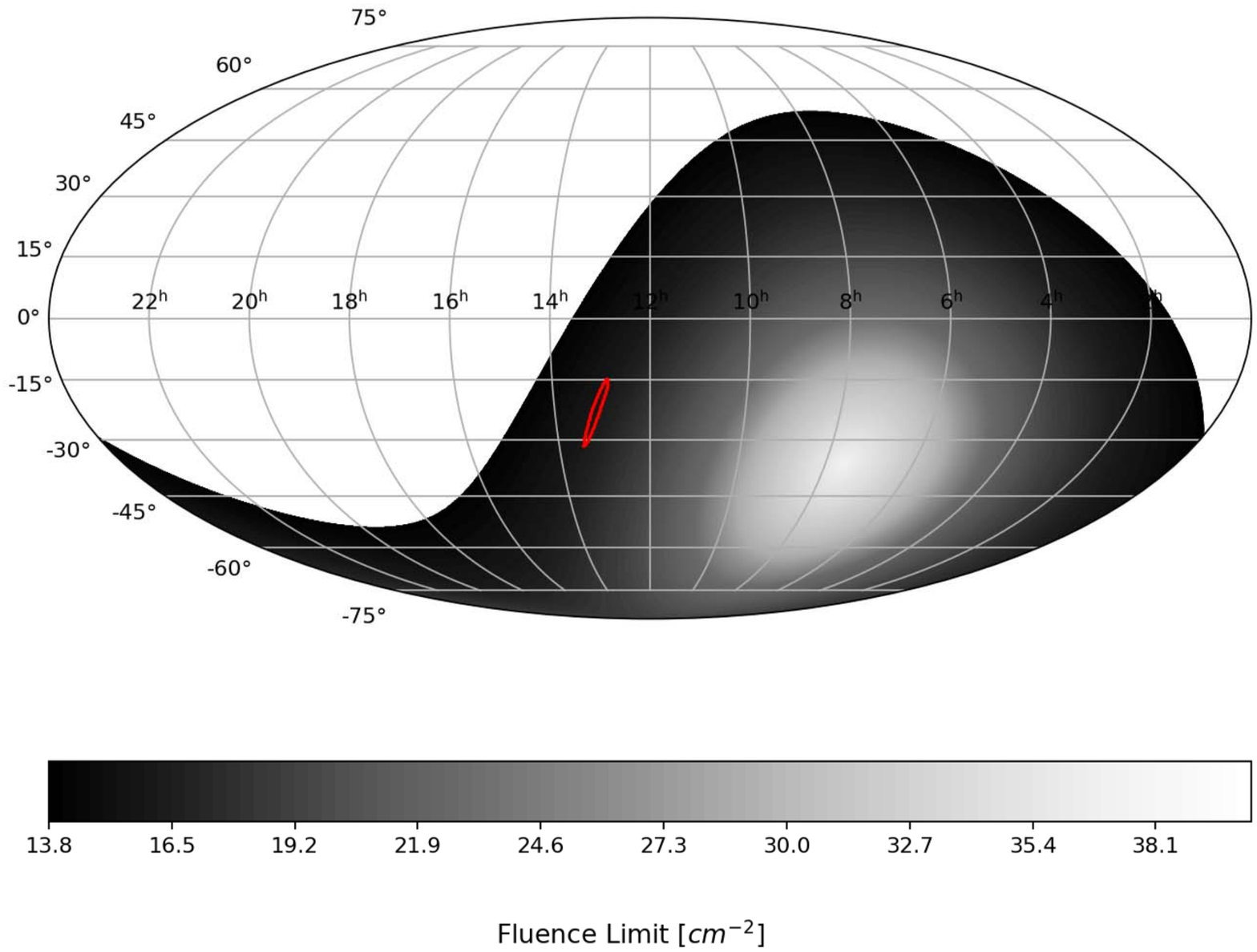}
\includegraphics[width=8cm]{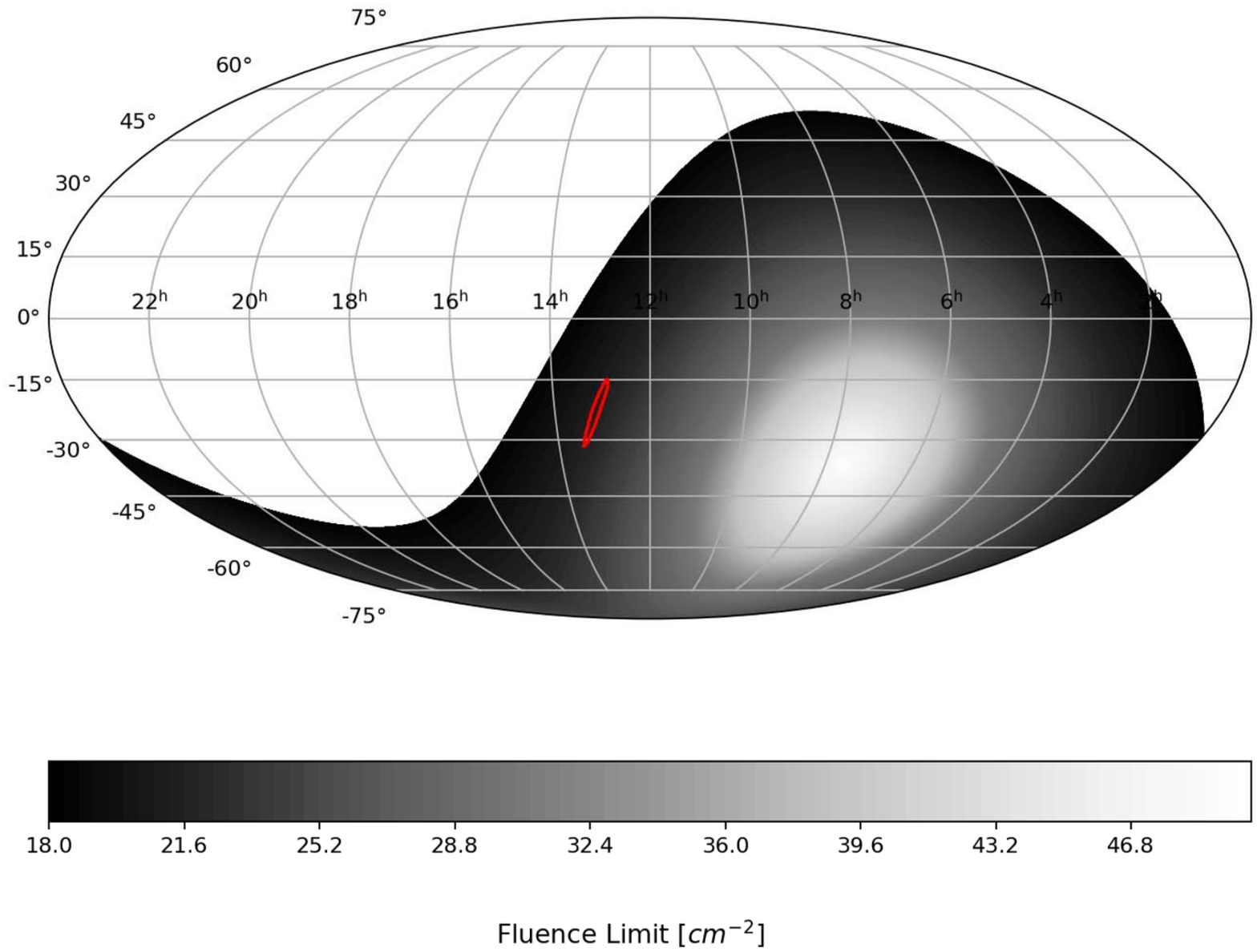}
\caption{The 90\% C.L. limit on fluence for neutrinos(left) and antineutrinos(right) in UPMU data set, overlaid with the 90\% C.L. contour for the location of GW170817 according to LIGO and Virgo released
 data (solid red line).
} \label{fig:UPMU_flu}
\end{figure}

\begin{figure}[hptb]
\begin{center}
\includegraphics[width=10cm]{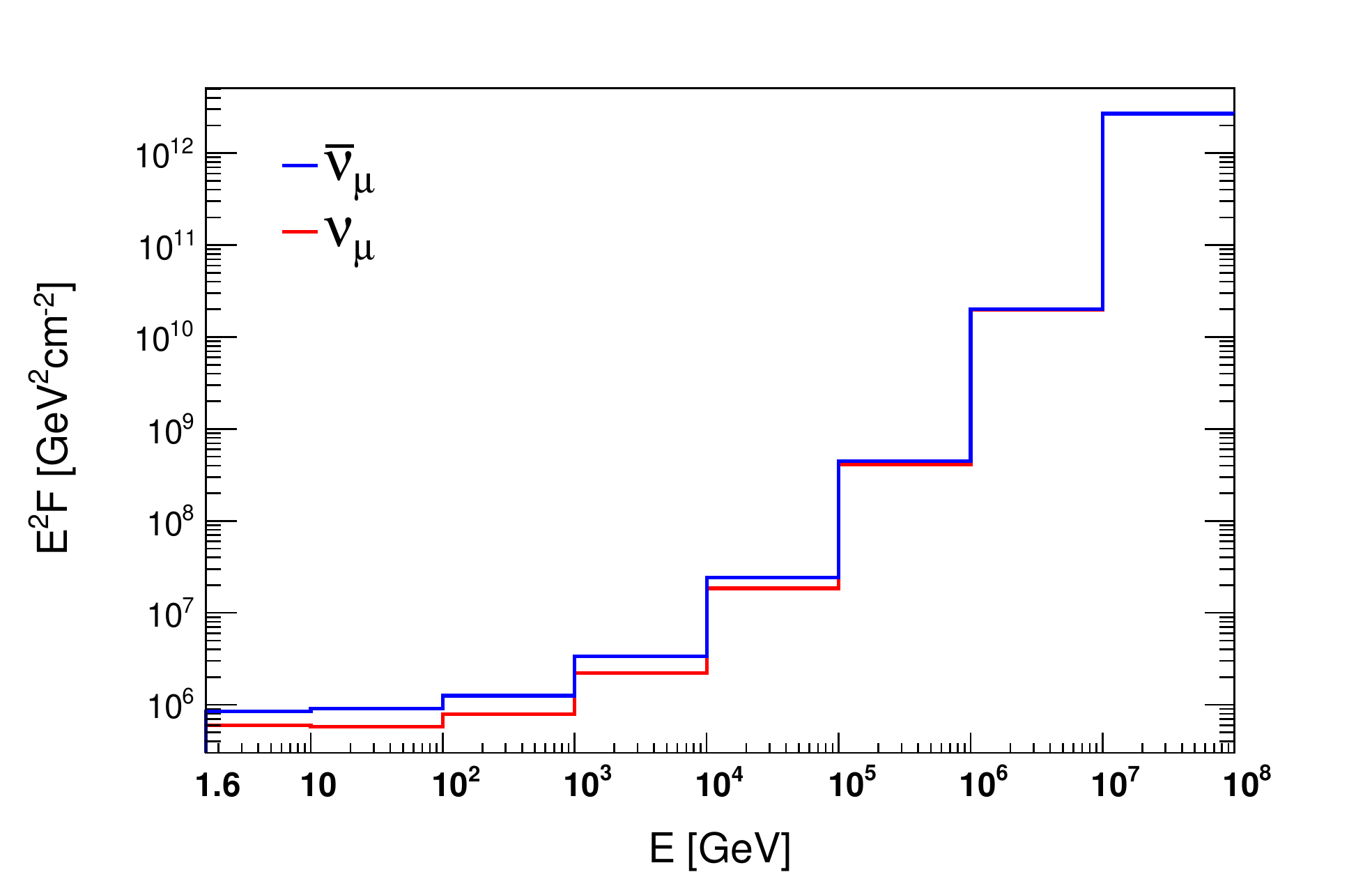}
\end{center}
\caption{The 90\% C.L. limits for the UPMU data set during a $\pm$500-s window, in the direction of NGC4993. Limits are calculated separately for each energy range, by assuming a spectrum with index of $-2$.}
\label{fig:UPMU_flu2}
\end{figure}

We also express the fluence limit which is calculated for monochromatic neutrino energy $E_{\nu}$.
The fluence limits at various energies are shown in Figure~\ref{fig:lowe_g}.
The results of fluence limits for FC+PC, UPMU, and low-energy data are summarized in Table~\ref{flu_table}. 
The UPMU upper limit fluence values range from (14--37) cm$^{-2}$ for neutrinos and from (18--50) cm$^{-2}$ for antineutrinos, depending on zenith angle from $90^{\circ}$ to $0^{\circ}$. To focus on the direction of NGC4993, UPMU limit is $16.0^{+0.7}_{-0.6}$ cm$^{-2}$ and $21.3^{+1.1}_{-0.8}$ cm$^{-2}$ for neutrinos and antineutrinos, while the error is calculated by a range of $\pm5^{\circ}$ around the zenith angle of NGC4993. We show the upper limit of neutrino fluence from UPMU events as a sky map in Figure~\ref{fig:UPMU_flu}. 
We note that the present study is sensitive to neutrinos between 1.6 GeV and 100 GeV, which is not covered in other searches~\cite{2017ApJ...850L..35A}. Our UPMU data may be compared or combined directly with that of other neutrino telescopes. We provide an UPMU fluency limit in Figure~\ref{fig:UPMU_flu2}.

Considering $d_{GW}$ as the distance from the detector to NGC4993, our upper limit on fluence of UPMU data can be converted into an upper limit on total radiated energy in neutrinos, by weighting by $4 \pi  d_{GW}^2$ in equation (\ref{eqn:UPMUfluence}).
The resulting upper limit on total energy is $E_{\nu}^{\mathrm{tot}} \sim $(1--6)$\times 10^{53}$ ergs for GW170817 assuming the luminosity distance of 40 Mpc. 

\begin{table}[hptb]
\begin{center}
\begin{tabular}{lc c c}\hline \hline
 & GW170817 $\Phi_{\nu} $(cm$^{-2}$) \\ \hline
 & {\bf from FC+PC only} & {\bf from UPMU only}\\
$\nu_{\mu}$       & $5.6 \times 10^4$  &   $16.0^{+0.7}_{-0.6}$\\
$\bar{\nu}_{\mu}$ &  $1.3 \times 10^5$ &  $21.3^{+1.1}_{-0.8}$\\
$\nu_e$           & $4.8 \times 10^4$ & -\\
$\bar{\nu}_e$     & $1.2 \times 10^5$ & -\\
\hline
 & {\bf from low-energy only} \\
 & flat spectrum & Fermi-Dirac with $E_{ave}$=20~MeV \\
$\bar{\nu}_e$     & 1.2 $\times 10^7$ & 6.6 $\times 10^7$ \\
$\nu_e$           & 1.0 $\times 10^9$ & 3.4 $\times 10^9$ \\
$\bar{\nu}_x$     & 7.5 $\times 10^9$ & 2.6 $\times 10^{10}$ \\
$\nu_x$           & 6.3 $\times 10^9$ & 2.1 $\times 10^{10}$ \\
\hline
\end{tabular}
\caption{Limits at 90\% C.L. on the fluence of neutrinos from GW170817 given a spectral index of $-2$ and a range of 100 MeV--10 GeV for FC+PC and 1.6 GeV--100~PeV for UPMU data. The error of UPMU limit is made with $\pm5^{\circ}$ range around zenith angle of NGC4993. Low-energy limits assume a flat spectrum as well as a Fermi-Dirac spectrum with $E_{average}$=20~MeV from 3.5~MeV to 100~MeV. $\nu_x$($\bar{\nu}_x$) represents $\nu_e$($\bar{\nu}_e$) and $\nu_{\mu}$($\bar{\nu}_{\mu}$).}
\label{flu_table}
\end{center}
\end{table}

\section{Conclusion}\label{sec:conclusion}
We made a coincidence search for neutrino signals with the gravitational wave, GW170817, produced by a binary neutron star merger in NGC4993, in the Super-Kamiokande detector in an energy range from 3.5~MeV to $\sim$100~PeV.
The analysis was performed within a time window of $\pm$500 s of GW170817 and 14 days after the neutron star merger.

In the high-energy data sample, three neutrino interaction categories are considered: FC, PC and UPMU.
No neutrino candidate was found in the $\pm$500-s window.
The numbers of candidates in a 14-day time window in the entire sky, as well as in a limited spatial region around NGC4993, are consistent with the expectation.

Low-energy neutrino events were also examined using the SRN and the solar neutrino data samples in the same window.
No neutrino candidate was found in the SRN and solar neutrino data samples in the $\pm$500-s window.
Two candidates were found in the SRN data sample in the 14-day search window, which is consistent with the estimated background rate.

Considering the observation of no significant neutrino signal associated with the GW170817 in SK, we calculated the neutrino fluence limits.
The obtained results give the most stringent limits for neutrino emission in the energy region below 100~GeV.

\begin{acknowledgments}
We gratefully acknowledge the cooperation of the Kamioka Mining and Smelting Company.
The Super-Kamiokande experiment has been built and operated from funding by the
Japanese Ministry of Education, Culture, Sports, Science and Technology, the U.S.
Department of Energy, and the U.S. National Science Foundation. Some of us have been
supported by funds from the National Research Foundation of Korea NRF-2009-0083526
(KNRC) funded by the Ministry of Science, ICT, and Future Planning, the European Union
H2020 RISE-GA641540-SKPLUS, the Japan Society for the Promotion of Science, the National
Natural Science Foundation of China under Grants No. 11235006, the National Science and
Engineering Research Council (NSERC) of Canada, the Scinet and Westgrid consortia of
Compute Canada, and the National Science Centre, Poland (2015/17/N/ST2/04064,
2015/18/E/ST2/00758).
\end{acknowledgments}


\begin{thebibliography}{}
\expandafter\ifx\csname natexlab\endcsname\relax\def\natexlab#1{#1}\fi

\bibitem[{{Abbott} {et~al.}(2017{\natexlab{a}}){Abbott}, {Abbott}, {Abbott},
  {Acernese}, {Ackley}, {Adams}, {Adams}, {Addesso}, {Adhikari}, {Adya}, \&
  et~al.}]{2017ApJ...848L..13A}
{Abbott}, B.~P., {Abbott}, R., {Abbott}, T.~D., {et~al.} 2017{\natexlab{a}},
  The Astrophysical Journal Letters, 848, L13

\bibitem[{{Abbott} {et~al.}(2017{\natexlab{b}}){Abbott}, {Abbott}, {Abbott},
  {Acernese}, {Ackley}, {Adams}, {Adams}, {Addesso}, {Adhikari}, {Adya}, \&
  et~al.}]{2017PhRvL.119p1101A}
---. 2017{\natexlab{b}}, Physical Review Letters, 119, 161101

\bibitem[{{Abbott} {et~al.}(2017{\natexlab{c}}){Abbott}, {Abbott}, {Abbott},
  {Acernese}, {Ackley}, {Adams}, {Adams}, {Addesso}, {Adhikari}, {Adya}, \&
  et~al.}]{2017ApJ...848L..12A}
---. 2017{\natexlab{c}}, The Astrophysical Journal Letters, 848, L12

\bibitem[{{Abe} {et~al.}(2014){Abe}, {Hayato}, {Iida}, {Iyogi}, {Kameda},
  {Kishimoto}, {Koshio}, {Marti}, {Miura}, {Moriyama}, {Nakahata}, {Nakano},
  {Nakayama}, {Obayashi}, {Sekiya}, {Shiozawa}, {Suzuki}, {Takeda}, {Takenaga},
  {Tanaka}, {Tomura}, {Ueno}, {Wendell}, {Yokozawa}, {Irvine}, {Kaji},
  {Kajita}, {Kaneyuki}, {Lee}, {Nishimura}, {Okumura}, {McLachlan}, {Labarga},
  {Kearns}, {Raaf}, {Stone}, {Sulak}, {Berkman}, {Tanaka}, {Tobayama},
  {Goldhaber}, {Bays}, {Carminati}, {Kropp}, {Mine}, {Renshaw}, {Smy}, {Sobel},
  {Ganezer}, {Hill}, {Keig}, {Jang}, {Kim}, {Lim}, {Hong}, {Akiri}, {Albert},
  {Himmel}, {Scholberg}, {Walter}, {Wongjirad}, {Ishizuka}, {Tasaka},
  {Learned}, {Matsuno}, {Smith}, {Hasegawa}, {Ishida}, {Ishii}, {Kobayashi},
  {Nakadaira}, {Nakamura}, {Nishikawa}, {Oyama}, {Sakashita}, {Sekiguchi},
  {Tsukamoto}, {Suzuki}, {Takeuchi}, {Huang}, {Ieki}, {Ikeda}, {Kikawa},
  {Kubo}, {Minamino}, {Murakami}, {Nakaya}, {Otani}, {Suzuki}, {Takahashi},
  {Fukuda}, {Choi}, {Itow}, {Mitsuka}, {Miyake}, {Mijakowski}, {Tacik},
  {Hignight}, {Imber}, {Jung}, {Taylor}, {Yanagisawa}, {Idehara}, {Ishino},
  {Kibayashi}, {Mori}, {Sakuda}, {Yamaguchi}, {Yano}, {Kuno}, {Kim}, {Yang},
  {Okazawa}, {Choi}, {Nishijima}, {Koshiba}, {Totsuka}, {Yokoyama}, {Martens},
  {Vagins}, {Martin}, {de Perio}, {Konaka}, {Wilking}, {Chen}, {Heng}, {Sui},
  {Yang}, {Zhang}, {Zhenwei}, {Connolly}, {Dziomba}, \&
  {Wilkes}}]{2014NIMPA.737..253A}
{Abe}, K., {Hayato}, Y., {Iida}, T., {et~al.} 2014, Nuclear Instruments and
  Methods in Physics Research A, 737, 253

\bibitem[{{Abe} {et~al.}(2016){Abe}, {Haga}, {Hayato}, {Ikeda}, {Iyogi},
  {Kameda}, {Kishimoto}, {Miura}, {Moriyama}, {Nakahata}, {Nakajima}, {Nakano},
  {Nakayama}, {Orii}, {Sekiya}, {Shiozawa}, {Takeda}, {Tanaka}, {Tasaka},
  {Tomura}, {Akutsu}, {Kajita}, {Kaneyuki}, {Nishimura}, {Richard}, {Okumura},
  {Labarga}, {Fernandez}, {Blaszczyk}, {Gustafson}, {Kachulis}, {Kearns},
  {Raaf}, {Stone}, {Sulak}, {Berkman}, {Nantais}, {Tobayama}, {Goldhaber},
  {Kropp}, {Mine}, {Weatherly}, {Smy}, {Sobel}, {Takhistov}, {Ganezer},
  {Hartfiel}, {Hill}, {Hong}, {Kim}, {Lim}, {Park}, {Himmel}, {Li},
  {O'Sullivan}, {Scholberg}, {Walter}, {Ishizuka}, {Nakamura}, {Jang}, {Choi},
  {Learned}, {Matsuno}, {Smith}, {Friend}, {Hasegawa}, {Ishida}, {Ishii},
  {Kobayashi}, {Nakadaira}, {Nakamura}, {Oyama}, {Sakashita}, {Sekiguchi},
  {Tsukamoto}, {Suzuki}, {Takeuchi}, {Yano}, {Cao}, {Hiraki}, {Hirota},
  {Huang}, {Jiang}, {Minamino}, {Nakaya}, {Patel}, {Wendell}, {Suzuki},
  {Fukuda}, {Itow}, {Suzuki}, {Mijakowski}, {Frankiewicz}, {Hignight}, {Imber},
  {Jung}, {Li}, {Palomino}, {Santucci}, {Wilking}, {Yanagisawa}, {Fukuda},
  {Ishino}, {Kayano}, {Kibayashi}, {Koshio}, {Mori}, {Sakuda}, {Xu}, {Kuno},
  {Tacik}, {Kim}, {Okazawa}, {Choi}, {Nishijima}, {Koshiba}, {Totsuka}, {Suda},
  {Yokoyama}, {Bronner}, {Calland}, {Hartz}, {Martens}, {Marti}, {Suzuki},
  {Vagins}, {Martin}, {Tanaka}, {Konaka}, {Chen}, {Wan}, {Zhang}, {Wilkes}, \&
  {Super-Kamiokande Collaboration}}]{2016ApJ...830L..11A}
{Abe}, K., {Haga}, K., {Hayato}, Y., {et~al.} 2016, The Astrophysical Journal
  Letters, 830, L11

\bibitem[{Abe {et~al.}(2016)Abe, Haga, Hayato, Ikeda, Iyogi, Kameda, Kishimoto,
  Marti, Miura, Moriyama, Nakahata, Nakajima, Nakayama, Orii, Sekiya, Shiozawa,
  Sonoda, Takeda, Tanaka, Takenaga, Tasaka, Tomura, Ueno, Yokozawa, Akutsu,
  Irvine, Kaji, Kajita, Kametani, Kaneyuki, Lee, Nishimura, McLachlan, Okumura,
  Richard, Labarga, Fernandez, Blaszczyk, Gustafson, Kachulis, Kearns, Raaf,
  Stone, Sulak, Berkman, Tobayama, Goldhaber, Bays, Carminati, Griskevich,
  Kropp, Mine, Renshaw, Smy, Sobel, Takhistov, Weatherly, Ganezer, Hartfiel,
  Hill, Keig, Hong, Kim, Lim, Park, Akiri, Albert, Himmel, Li, O'Sullivan,
  Scholberg, Walter, Wongjirad, Ishizuka, Nakamura, Jang, Choi, Learned,
  Matsuno, Smith, Friend, Hasegawa, Ishida, Ishii, Kobayashi, Nakadaira,
  Nakamura, Nishikawa, Oyama, Sakashita, Sekiguchi, Tsukamoto, Nakano, Suzuki,
  Takeuchi, Yano, Cao, Hayashino, Hiraki, Hirota, Huang, Ieki, Jiang, Kikawa,
  Minamino, Murakami, Nakaya, Patel, Suzuki, Takahashi, Wendell, Fukuda, Itow,
  Mitsuka, Muto, Suzuki, Mijakowski, Frankiewicz, Hignight, Imber, Jung, Li,
  Palomino, Santucci, Taylor, Vilela, Wilking, Yanagisawa, Fukuda, Ishino,
  Kayano, Kibayashi, Koshio, Mori, Sakuda, Takeuchi, Yamaguchi, Kuno, Tacik,
  Kim, Okazawa, Choi, Ito, Nishijima, Koshiba, Totsuka, Suda, Yokoyama,
  Bronner, Calland, Hartz, Martens, Obayashi, Suzuki, Vagins, Nantais, Martin,
  de~Perio, Tanaka, Konaka, Chen, Sui, Wan, Yang, Zhang, Zhang, Connolly,
  Dziomba, \& Wilkes}]{sk4sol}
Abe, K., Haga, Y., Hayato, Y., {et~al.} 2016, Physical Review D, 94, 052010

\bibitem[{Abe {et~al.}(2017)Abe, Bronner, Pronost, Hayato, Ikeda, Iyogi,
  Kameda, Kato, Kishimoto, Marti, Miura, Moriyama, Nakahata, Nakano, Nakayama,
  Okajima, Orii, Sekiya, Shiozawa, Sonoda, Takeda, Takenaka, Tanaka, Tasaka,
  Tomura, Akutsu, Kajita, Kaneyuki, Nishimura, Okumura, Tsui, Labarga,
  Fernandez, d.~M.~Blaszczyk, Gustafson, Kachulis, Kearns, Raaf, Stone, Sulak,
  Berkman, Tobayama, Goldhaber, Elnimr, Kropp, Mine, Locke, Weatherly, Smy,
  Sobel, Takhistov, Ganezer, Hill, Kim, Lim, Park, Himmel, Li, O'Sullivan,
  Scholberg, Walter, Ishizuka, Nakamura, Jang, Choi, Learned, Matsuno, Smith,
  Amey, Litchfield, Ma, Uchida, Wascko, Cao, Friend, Hasegawa, Ishida, Ishii,
  Kobayashi, Nakadaira, Nakamura, Oyama, Sakashita, Sekiguchi, Tsukamoto, Abe,
  Hasegawa, Suzuki, Takeuchi, Yano, Cao, Hayashino, Hiraki, Hirota, Huang,
  Jiang, Minamino, Nakamura, Nakaya, Quilain, Patel, Wendell, Anthony,
  McCauley, Pritchard, Fukuda, Itow, Murase, Muto, Mijakowski, Frankiewicz,
  Jung, Li, Palomino, Santucci, Vilela, Wilking, Yanagisawa, Ito, Fukuda,
  Ishino, Kibayashi, Koshio, Nagata, Sakuda, Xu, Kuno, Wark, Lodovico,
  Richards, Tacik, Kim, Cole, Thompson, Okazawa, Choi, Ito, Nishijima, Koshiba,
  Totsuka, Suda, Yokoyama, Calland, Hartz, Martens, Simpson, Suzuki, Vagins,
  Hamabe, Kuze, Yoshida, Ishitsuka, Martin, Nantais, Tanaka, Konaka, Chen, Wan,
  Zhang, Minamino, Wilkes, \& Collaboration}]{erin_icecube}
Abe, K., Bronner, C., Pronost, G., {et~al.} 2017, The Astrophysical Journal,
  850, 166

\bibitem[{{Albert} {et~al.}(2017){Albert}, {Andr{\'e}}, {Anghinolfi}, {Ardid},
  {Aubert}, {Aublin}, {Avgitas}, {Baret}, {Barrios-Mart{\'{\i}}}, {Basa}, \&
  et~al.}]{2017ApJ...850L..35A}
{Albert}, A., {Andr{\'e}}, M., {Anghinolfi}, M., {et~al.} 2017, The
  Astrophysical Journal Letters, 850, L35

\bibitem[{Ashie {et~al.}(2005)Ashie, Hosaka, Ishihara, Itow, Kameda, Koshio,
  Minamino, Mitsuda, Miura, Moriyama, Nakahata, Namba, Nambu, Obayashi,
  Shiozawa, Suzuki, Takeuchi, Taki, Yamada, Ishitsuka, Kajita, Kaneyuki,
  Nakayama, Okada, Okumura, Saji, Takenaga, Clark, Desai, Kearns, Likhoded,
  Stone, Sulak, Wang, Goldhaber, Casper, Cravens, Gajewski, Kropp, Liu, Mine,
  Smy, Sobel, Sterner, Vagins, Ganezer, Hill, Keig, Jang, Kim, Lim, Scholberg,
  Walter, Ellsworth, Tasaka, Guillian, Kibayashi, Learned, Matsuno, Takemori,
  Messier, Hayato, Ichikawa, Ishida, Ishii, Iwashita, Kobayashi, Maruyama,
  Nakamura, Nitta, Oyama, Sakuda, Totsuka, Suzuki, Hasegawa, Hayashi, Kato,
  Maesaka, Morita, Nakaya, Nishikawa, Sasaki, Ueda, Yamamoto, Haines, Dazeley,
  Hatakeyama, Svoboda, Blaufuss, Goodman, Sullivan, Turcan, Habig, Fukuda,
  Jung, Kato, Kobayashi, Malek, Mauger, McGrew, Sarrat, Sharkey, Yanagisawa,
  Toshito, Miyano, Tamura, Ishii, Kuno, Yoshida, Kim, Yoo, Okazawa, Ishizuka,
  Choi, Seo, Gando, Hasegawa, Inoue, Shirai, Suzuki, Koshiba, Nakajima,
  Nishijima, Harada, Ishino, Watanabe, Kielczewska, Zalipska, Berns, Gran,
  Shiraishi, Stachyra, Washburn, \& Wilkes}]{ashie05}
Ashie, Y., Hosaka, J., Ishihara, K., {et~al.} 2005, Physical Review D, 71,
  112005

\bibitem[{{Bays} {et~al.}(2012){Bays}, {Iida}, {Abe}, {Hayato}, {Iyogi},
  {Kameda}, {Koshio}, {Marti}, {Miura}, {Moriyama}, {Nakahata}, {Nakayama},
  {Obayashi}, {Sekiya}, {Shiozawa}, {Suzuki}, {Takeda}, {Takenaga}, {Ueno},
  {Ueshima}, {Yamada}, {Yokozawa}, {Kaji}, {Kajita}, {Kaneyuki}, {McLachlan},
  {Okumura}, {Lee}, {Martens}, {Vagins}, {Labarga}, {Kearns}, {Litos}, {Raaf},
  {Stone}, {Sulak}, {Kropp}, {Mine}, {Regis}, {Renshaw}, {Smy}, {Sobel},
  {Ganezer}, {Hill}, {Keig}, {Cho}, {Jang}, {Kim}, {Lim}, {Albert},
  {Scholberg}, {Walter}, {Wendell}, {Wongjirad}, {Ishizuka}, {Tasaka},
  {Learned}, {Matsuno}, {Smith}, {Hasegawa}, {Ishida}, {Ishii}, {Kobayashi},
  {Nakadaira}, {Nakamura}, {Nishikawa}, {Oyama}, {Sakashita}, {Sekiguchi},
  {Tsukamoto}, {Suzuki}, {Takeuchi}, {Ikeda}, {Matsuoka}, {Minamino},
  {Murakami}, {Nakaya}, {Fukuda}, {Itow}, {Mitsuka}, {Miyake}, {Tanaka},
  {Hignight}, {Imber}, {Jung}, {Taylor}, {Yanagisawa}, {Kibayashi}, {Ishino},
  {Mino}, {Sakuda}, {Mori}, {Toyota}, {Kuno}, {Kim}, {Yang}, {Okazawa}, {Choi},
  {Nishijima}, {Koshiba}, {Totsuka}, {Yokoyama}, {Heng}, {Chen}, {Zhang},
  {Yang}, {Mijakowski}, {Connolly}, {Dziomba}, \& {Wilkes}}]{sksrn}
{Bays}, K., {Iida}, T., {Abe}, K., {et~al.} 2012, Physical Review D, 85, 052007

\bibitem[{{Fukuda} {et~al.}(2003){Fukuda}, {Fukuda}, {Hayakawa}, {Ichihara},
  {Ishitsuka}, {Itow}, {Kajita}, {Kameda}, {Kaneyuki}, {Kasuga}, {Kobayashi},
  {Kobayashi}, {Koshio}, {Miura}, {Moriyama}, {Nakahata}, {Nakayama}, {Namba},
  {Obayashi}, {Okada}, {Oketa}, {Okumura}, {Oyabu}, {Sakurai}, {Shiozawa},
  {Suzuki}, {Takeuchi}, {Toshito}, {Totsuka}, {Yamada}, {Desai}, {Earl},
  {Hong}, {Kearns}, {Masuzawa}, {Messier}, {Stone}, {Sulak}, {Walter}, {Wang},
  {Scholberg}, {Barszczak}, {Casper}, {Liu}, {Gajewski}, {Halverson}, {Hsu},
  {Kropp}, {Mine}, {Price}, {Reines}, {Smy}, {Sobel}, {Vagins}, {Ganezer},
  {Keig}, {Ellsworth}, {Tasaka}, {Flanagan}, {Kibayashi}, {Learned}, {Matsuno},
  {Stenger}, {Hayato}, {Ishii}, {Ichikawa}, {Kanzaki}, {Kobayashi}, {Maruyama},
  {Nakamura}, {Oyama}, {Sakai}, {Sakuda}, {Sasaki}, {Echigo}, {Iwashita},
  {Kohama}, {Suzuki}, {Hasegawa}, {Inagaki}, {Kato}, {Maesaka}, {Nakaya},
  {Nishikawa}, {Yamamoto}, {Haines}, {Kim}, {Sanford}, {Svoboda}, {Blaufuss},
  {Chen}, {Conner}, {Goodman}, {Guillian}, {Sullivan}, {Turcan}, {Habig},
  {Ackerman}, {Goebel}, {Hill}, {Jung}, {Kato}, {Kerr}, {Malek}, {Martens},
  {Mauger}, {McGrew}, {Sharkey}, {Viren}, {Yanagisawa}, {Doki}, {Inaba}, {Ito},
  {Kirisawa}, {Kitaguchi}, {Mitsuda}, {Miyano}, {Saji}, {Takahata},
  {Takahashi}, {Higuchi}, {Kajiyama}, {Kusano}, {Nagashima}, {Nitta}, {Takita},
  {Yamaguchi}, {Yoshida}, {Kim}, {Kim}, {Yoo}, {Okazawa}, {Etoh}, {Fujita},
  {Gando}, {Hasegawa}, {Hasegawa}, {Hatakeyama}, {Inoue}, {Ishihara},
  {Iwamoto}, {Koga}, {Nishiyama}, {Ogawa}, {Shirai}, {Suzuki}, {Takayama},
  {Tsushima}, {Koshiba}, {Ichikawa}, {Hashimoto}, {Hatakeyama}, {Koike},
  {Horiuchi}, {Nemoto}, {Nishijima}, {Takeda}, {Fujiyasu}, {Futagami},
  {Ishino}, {Kanaya}, {Morii}, {Nishihama}, {Nishimura}, {Suzuki}, {Watanabe},
  {Kielczewska}, {Golebiewska}, {Berns}, {Boyd}, {Doyle}, {George}, {Stachyra},
  {Wai}, {Wilkes}, {Young}, {Kobayashi}, \& {Super-Kamiokande
  Collaboration}}]{2003NIMPA.501..418F}
{Fukuda}, S., {Fukuda}, Y., {Hayakawa}, T., {et~al.} 2003, Nuclear Instruments
  and Methods in Physics Research A, 501, 418

\bibitem[{Gaisser {et~al.}(1995)Gaisser, Halzen, \& Stanev}]{gaisser94}
Gaisser, T.~K., Halzen, F., \& Stanev, T. 1995, Physics Reports, 258, 173

\bibitem[{Hayato(2009)}]{hayato09}
Hayato, Y. 2009, Acta Physica Polonica B, B40, 2477

\bibitem[{{Kyutoku} \& {Kashiyama}(2017)}]{1710.05922}
{Kyutoku}, K., \& {Kashiyama}, K. 2017, arXiv:1710.05922

\bibitem[{{Nakahata} {et~al.}(1999){Nakahata}, {Fukuda}, {Hayakawa},
  {Ichihara}, {Inoue}, {Ishihara}, {Ishino}, {Itow}, {Kajita}, {Kameda},
  {Kasuga}, {Kobayashi}, {Kobayashi}, {Koshio}, {Martens}, {Miura}, {Nakayama},
  {Okada}, {Okumura}, {Sakurai}, {Shiozawa}, {Suzuki}, {Takeuchi}, {Totsuka},
  {Yamada}, {Earl}, {Habig}, {Kearns}, {Messier}, {Scholberg}, {Stone},
  {Sulak}, {Walter}, {Goldhaber}, {Barszczak}, {Casper}, {Gajewski},
  {Halverson}, {Hsu}, {Kropp}, {Price}, {Reines}, {Smy}, {Sobel}, {Vagins},
  {Ganezer}, {Keig}, {Ellsworth}, {Tasaka}, {Flanagan}, {Kibayashi}, {Learned},
  {Matsuno}, {Stenger}, {Takemori}, {Ishii}, {Kanzaki}, {Kobayashi}, {Mine},
  {Nakamura}, {Nishikawa}, {Oyama}, {Sakai}, {Sakuda}, {Sasaki}, {Echigo},
  {Kohama}, {Suzuki}, {Haines}, {Blaufuss}, {Kim}, {Sanford}, {Svoboda},
  {Chen}, {Conner}, {Goodman}, {Sullivan}, {Hill}, {Jung}, {Mauger}, {McGrew},
  {Sharkey}, {Viren}, {Yanagisawa}, {Doki}, {Miyano}, {Okazawa}, {Saji},
  {Takahata}, {Nagashima}, {Takita}, {Yamaguchi}, {Yoshida}, {Kim}, {Etoh},
  {Fujita}, {Hasegawa}, {Hasegawa}, {Hatakeyama}, {Iwamoto}, {Koga},
  {Maruyama}, {Ogawa}, {Shirai}, {Suzuki}, {Tsushima}, {Koshiba}, {Nemoto},
  {Nishijima}, {Futagami}, {Hayato}, {Kanaya}, {Kaneyuki}, {Watanabe},
  {Kielczewska}, {Doyle}, {George}, {Stachyra}, {Wai}, {Wilkes}, {Young}, \&
  {Kobayashi}}]{1999NIMPA.421..113N}
{Nakahata}, M., {Fukuda}, Y., {Hayakawa}, T., {et~al.} 1999, Nuclear
  Instruments and Methods in Physics Research A, 421, 113

\bibitem[{{Sekiguchi} {et~al.}(2011){Sekiguchi}, {Kiuchi}, {Kyutoku}, \&
  {Shibata}}]{2011PhRvL.107e1102S}
{Sekiguchi}, Y., {Kiuchi}, K., {Kyutoku}, K., \& {Shibata}, M. 2011, Physical
  Review Letters, 107, 051102

\bibitem[{{Swanson} {et~al.}(2006){Swanson}, {Abe}, {Hosaka}, {Iida},
  {Ishihara}, {Kameda}, {Koshio}, {Minamino}, {Mitsuda}, {Miura}, {Moriyama},
  {Nakahata}, {Obayashi}, {Ogawa}, {Shiozawa}, {Suzuki}, {Takeda}, {Takeuchi},
  {Ueshima}, {Higuchi}, {Ishihara}, {Ishitsuka}, {Kajita}, {Kaneyuki},
  {Mitsuka}, {Nakayama}, {Nishino}, {Okada}, {Okumura}, {Saji}, {Takenaga},
  {Clark}, {Desai}, {Dufour}, {Kearns}, {Likhoded}, {Litos}, {Raaf}, {Stone},
  {Sulak}, {Wang}, {Goldhaber}, {Casper}, {Cravens}, {Dunmore}, {Kropp}, {Liu},
  {Mine}, {Regis}, {Smy}, {Sobel}, {Vagins}, {Ganezer}, {Hill}, {Keig}, {Jang},
  {Kim}, {Lim}, {Scholberg}, {Tanimoto}, {Walter}, {Wendell}, {Ellsworth},
  {Tasaka}, {Guillian}, {Learned}, {Matsuno}, {Messier}, {Hayato}, {Ichikawa},
  {Ishida}, {Ishii}, {Iwashita}, {Kobayashi}, {Nakadaira}, {Nakamura}, {Nitta},
  {Oyama}, {Totsuka}, {Suzuki}, {Hasegawa}, {Hiraide}, {Kato}, {Maesaka},
  {Nakaya}, {Nishikawa}, {Sasaki}, {Sato}, {Yamamoto}, {Yokoyama}, {Haines},
  {Dazeley}, {Hatakeyama}, {Svoboda}, {Sullivan}, {Turcan}, {Cooley}, {Mahn},
  {Habig}, {Fukuda}, {Sato}, {Itow}, {Koike}, {Jung}, {Kato}, {Kobayashi},
  {Malek}, {McGrew}, {Sarrat}, {Terri}, {Yanagisawa}, {Tamura}, {Sakuda},
  {Sugihara}, {Kuno}, {Yoshida}, {Kim}, {Yang}, {Yoo}, {Ishizuka}, {Okazawa},
  {Choi}, {Seo}, {Gando}, {Hasegawa}, {Inoue}, {Ishii}, {Nishijima}, {Ishino},
  {Watanabe}, {Koshiba}, {Kielczewska}, {Zalipska}, {Berns}, {Gran},
  {Shiraishi}, {Stachyra}, {Thrane}, {Washburn}, {Wilkes}, \& {Super-KAMIOKANDE
  Collaboration}}]{swanson06}
{Swanson}, M.~E.~C., {Abe}, K., {Hosaka}, J., {et~al.} 2006, The Astrophysical
  Journal, 652, 206

\bibitem[{{Thrane} {et~al.}(2009){Thrane}, {Abe}, {Hayato}, {Iida}, {Ikeda},
  {Kameda}, {Kobayashi}, {Koshio}, {Miura}, {Moriyama}, {Nakahata}, {Nakayama},
  {Obayashi}, {Ogawa}, {Sekiya}, {Shiozawa}, {Suzuki}, {Takeda}, {Takenaga},
  {Takeuchi}, {Ueno}, {Ueshima}, {Watanabe}, {Yamada}, {Vagins}, {Hazama},
  {Higuchi}, {Ishihara}, {Kajita}, {Kaneyuki}, {Mitsuka}, {Nishino}, {Okumura},
  {Tanimoto}, {Dufour}, {Kearns}, {Litos}, {Raaf}, {Stone}, {Sulak},
  {Goldhaber}, {Bays}, {Casper}, {Cravens}, {Kropp}, {Mine}, {Regis}, {Smy},
  {Sobel}, {Ganezer}, {Hill}, {Keig}, {Jang}, {Jeong}, {Kim}, {Lim}, {Fechner},
  {Scholberg}, {Walter}, {Wendell}, {Tasaka}, {Learned}, {Matsuno}, {Watanabe},
  {Hasegawa}, {Ishida}, {Ishii}, {Kobayashi}, {Nakadaira}, {Nakamura},
  {Nishikawa}, {Oyama}, {Sakashita}, {Sekiguchi}, {Tsukamoto}, {Suzuki},
  {Ichikawa}, {Minamino}, {Nakaya}, {Yokoyama}, {Dazeley}, {Svoboda}, {Habig},
  {Fukuda}, {Itow}, {Tanaka}, {Jung}, {Lopez}, {McGrew}, {Yanagisawa},
  {Tamura}, {Idehara}, {Ishino}, {Kibayashi}, {Sakuda}, {Kuno}, {Yoshida},
  {Kim}, {Yang}, {Ishizuka}, {Okazawa}, {Choi}, {Seo}, {Furuse}, {Nishijima},
  {Yokosawa}, {Koshiba}, {Totsuka}, {Chen}, {Gong}, {Heng}, {Xue}, {Yang},
  {Zhang}, {Kielczewska}, {Mijakowski}, {Connolly}, {Dziomba}, \&
  {Wilkes}}]{thrane09}
{Thrane}, E., {Abe}, K., {Hayato}, Y., {et~al.} 2009, The Astrophysical
  Journal, 704, 503

\bibitem[{{Waxman} \& {Bahcall}(1997)}]{1997PhRvL..78.2292W}
{Waxman}, E., \& {Bahcall}, J. 1997, Physical Review Letters, 78, 2292

\bibitem[{{Zhang} {et~al.}(2015){Zhang}, {Abe}, {Hayato}, {Iyogi}, {Kameda},
  {Kishimoto}, {Miura}, {Moriyama}, {Nakahata}, {Nakano}, {Nakayama}, {Sekiya},
  {Shiozawa}, {Suzuki}, {Takeda}, {Takenaga}, {Tomura}, {Ueno}, {Yokozawa},
  {Wendell}, {Kaji}, {Kajita}, {Kaneyuki}, {Lee}, {Nishimura}, {Okumura},
  {McLachlan}, {Labarga}, {Barkman}, {Tanaka}, {Tobayama}, {Kearns}, {Raaf},
  {Stone}, {Sulak}, {Goldhaber}, {Bays}, {Carminati}, {Kropp}, {Mine},
  {Renshaw}, {Smy}, {Sobel}, {Ganezer}, {Hill}, {Keig}, {Jang}, {Kim}, {Lim},
  {Akiri}, {Scholberg}, {Walter}, {Wongjirad}, {Ishizuka}, {Tasaka}, {Learned},
  {Matsuno}, {Smith}, {Hasegawa}, {Ishida}, {Ishii}, {Kobayashi}, {Nakadaira},
  {Nakamura}, {Nishikawa}, {Oyama}, {Sakashita}, {Sekiguchi}, {Tsukamoto},
  {Suzuki}, {Takeuchi}, {Ieki}, {Ikeda}, {Kikawa}, {Huang}, {Minamino},
  {Murakami}, {Nakaya}, {Suzuki}, {Takahashi}, {Fukuda}, {Choi}, {Itow},
  {Mitsuka}, {Mijakowski}, {Hignight}, {Imber}, {Jung}, {Taylor}, {Yanagisawa},
  {Ishino}, {Kibayashi}, {Koshio}, {Mori}, {Sakuda}, {Yamaguchi}, {Yano},
  {Kuno}, {Tacik}, {Kim}, {Okazawa}, {Choi}, {Nishijima}, {Koshiba}, {Totsuka},
  {Yokoyama}, {Martens}, {Marti}, {Vagins}, {Martin}, {dePerio}, {Konaka},
  {Wilking}, {Chen}, {Sui}, {Yang}, {Zhang}, {Connolly}, {Dziomba}, \&
  {Wilkes}}]{ntag}
{Zhang}, H., {Abe}, K., {Hayato}, Y., {et~al.} 2015, Astroparticle Physics, 60,
  41

\end{thebibliography}

\end{document}